\documentclass[useAMS,usenatbib]{mn2e}
\usepackage{graphicx}
\title[BH masses in ULXs]{Low metallicity natal environments and black hole 
masses in Ultraluminous X-ray Sources}
\author[L. Zampieri and T. P. Roberts]{L. Zampieri$^1$ and T. P. Roberts$^2$ \\
$^1$ INAF-Osservatorio Astronomico di Padova, Vicolo dell'Osservatorio 5,
I-35122 Padova, Italy \\
$^2$ Department of Physics, Durham University, South Road, Durham DH1 3LE, UK
}
\def\la{\mathrel{\hbox{\rlap{\hbox{\lower4pt\hbox{$\sim$}}}{\raise2pt\hbox{$<$}}}}}
\def\ga{\mathrel{\hbox{\rlap{\hbox{\lower4pt\hbox{$\sim$}}}{\raise2pt\hbox{$>$}}}}}
\begin{document}

\date{Accepted ... Received ...; in original form ...}
%
%\pagerange{\pageref{firstpage}--\pageref{lastpage}} \pubyear{2002}
%
\maketitle
\label{firstpage}

%%%%%%%%%%%%%%%%%%%%%%%%%%%%%%%%%%%%%%%%%%%%%%%%%%%%%%%%%%%%%%%%%%%%%%%%%%%%%
%
%
%
%
\begin{abstract}
We review the available estimates of the masses of the compact object
in Ultraluminous X-ray Sources (ULXs) and critically reconsider the
stellar-mass versus intermediate-mass black hole
interpretations. Black holes of several hundreds to thousands of
$M_\odot$ are not required for the majority of ULXs, although they
might be present in the handful of known hyper-luminous ($\sim
10^{41}$ erg s$^{-1}$) objects and/or some sources showing timing
features in their power density spectra. At the same time, however,
stellar mass BHs may be quite a reasonable explanation for ULXs below
$\sim 10^{40}$ erg s$^{-1}$, but they need super-Eddington accretion
and some suitable dependence of the beaming factor on the accretion
rate in order to account for ULXs above this (isotropic) luminosity.
%The observational
%limits from current data, particularly those coming from X-ray
%spectroscopy, are consistent with black holes that are somewhat bigger
%than the known stellar-mass black holes in our Galaxy.  
We investigate in detail a 'third way' in which a proportion of ULXs contain 
$\approx 30-90 M_\odot$ black holes formed in a low metallicity environment
and accreting in a slightly critical regime and find that it can
consistently account for the properties of bright ULXs.
Surveys of ULX locations looking for a statistically
meaningful relationship between ULX position, average 
luminosity and local metallicity will provide a definitive test of our proposal.
\end{abstract}
\begin{keywords}
accretion, accretion discs --- black hole physics --- X-rays: binaries
\end{keywords}

%%%%%%%%%%%%%%%%%%%%%%%%%%%%%%%%%%%%%%%%%%%%%%%%%%%%%%%%%%%%%%%%%%%%%%%%%%%%%
%
%
%
%
\section{Introduction}
\label{sec1}

When, at the beginning of the '80s, point-like, off-nuclear X-ray
sources were first detected in the field of nearby galaxies (see,
e.g., \citealt{b17,b19}), it was immediately recognised that the
luminosity of a subset of these objects was unusually large.  If
physically associated with their host galaxies, these sources had an
isotropic luminosity well in excess of the Eddington limit for
spherical accretion onto a $10 M_{\odot}$ compact object.  These
apparently Super-Eddington sources, later called UltraLuminous X-ray
sources (ULXs), were first noticed in {\it Einstein\/} data
(\citealt{lvs83,h84,b19}).

Nowadays well in excess of 150 candidate ULXs have been detected and
catalogued by a variety of X-ray observatories (see e.g.
\citealt{Rob00,b12,b63,b41}).  A small fraction of ULXs
are now known to be X-ray luminous interacting supernovae such as
those described by \cite{Immler07}.  A much larger fraction have
subsequently been identified with background AGNs ($\sim$25\%;
\citealt{b63}; see also \citealt{b23b,mas03,wong08}).  This background
contamination is stronger in ellipticals, where it accounts for $\sim
44\%$ of all the ULXs, than in spirals ($\sim$14\%;
\citealt{b63}). However, the majority of these puzzling
Super-Eddington sources constitute a very interesting class, that
remains yet to be fully understood.

The recent detection of a $\sim$62 day modulation in the light curve
of M 82 X-1 has been interpreted as the orbital period of the system
(\citealt{k06a,k06b,fk07}).  A periodic modulation
(12.5 hrs) has also recently been detected in another ULX (NGC 3379;
\citealt{f06}). These results support the notion that ULXs are
X-ray binary systems, where mass transferred from a donor star falls
onto a compact object via an X-ray luminous accretion disc.  The high
X-ray luminosities of ULXs suggest this compact object is most likely a
black hole (BH).

The X-ray spectral properties of ULXs show similarities with those of
X-ray binaries (XRBs) in our Galaxy (e.g. \citealt{b23}).  In many
cases the spectrum can be well reproduced by a multicolour disc (MCD)
blackbody, representing emission from an accretion disc, plus a
power-law continuum (PL), with the latter nominally representing
emission from a Compton-scattering corona.  This is the same canonical
model employed to describe the spectra of Galactic black hole X-ray
binaries (cf. \citealt{MR06}).  Interestingly, the derived temperature
of the MCD component in ULXs is often much lower than that observed in
XRBs (e.g. \citealt{b44,b44b,b21}). However, for some of the brightest
ULXs, a possible curvature above 2-3 keV has been reported and equally
acceptable fits of their spectra may be obtained with (physically)
different models, that suggest the presence of hitherto unusual
features such as an optically thick corona, a fast ionised outflow or
a slim disc (e.g. \citealt{b61,b27,m07}).

Fast X-ray variability can also reveal much about accretion-powered
sources.  Although most ULXs show little variability on timescales of
seconds to hours (\citealt{b63,RWWG04}), the recent detections of
quasi periodic oscillations in the power density spectra of M 82 X-1
and NGC 5408 X-1 has shed new light on the timescales in the inner
accretion disc of these systems (\citealt{b62a,ft04,b49b,b62,cas08}).

Stellar optical counterparts have been discovered to be associated
with a number of ULXs
(\citealt{Rob01,GRKL02,b40,b40a,b30a,b73,k05,b49,sor05,pak06}), although
only some of them have been associated with stellar objects of known
spectral type (e.g. \citealt{b40,b40a,b30a,b49}).  In almost all
cases, the counterparts appear to be hosted in young stellar
environments (e.g. \citealt{ram06,pak06,liu07}) and have properties
consistent with those of young, massive stars. However, some ULXs
appear to be associated to older stellar populations and at least one
possible later-type stellar counterpart is now known
(\citealt{FK08,RLG08}).  Some ULXs are also associated with very
extended optical emission nebulae, that may provide important
information on the energetics and lifetime of these systems
(\citealt{b55,RGWW04}).  These nebulae are also beginning to be
detected as extended radio sources (\citealt{MMN05,Lang07}).

The stellar environment of ULXs can also provide interesting
constraints on the properties of ULX binary systems (ULXBs).  Several
ULXs are located in groups or clusters of OB stars. Isochrone fitting
of the cluster colour-magnitude diagram has been attempted and
provides cluster ages of tens of millions of years, although there is
some disagreement among different authors
(\citealt{ram06,pak06,liu07,grise08}). These analyses translate into
upper limits for the donor masses in ULXBs, assuming that they are
coeval to their parent OB association.
%Using the same assumptions, such characteristic ages also provide lower limits for
%the mass of the BH progenitor, as it must have already evolved off the
%main sequence to form the compact object.
Typical values of these mass limits are in the range $\sim 10-20
M_\odot$.  Comparison of stellar evolutionary tracks of ULXs with the
photometric properties of their optical counterparts on the
colour-magnitude diagram may also be used to constrain the masses of
their donor stars (e.g. \citealt{sor05,cop05,cop07}). If accurate
photometry is available, this approach may also provide interesting
clues to the BH mass, once binary evolution and X-ray irradiation
effects are taken into account (\citealt{pz08}).

ULXs play a fundamental role in the framework of X-ray source
populations in nearby galaxies and can be detected and studied at
larger distances than more 'normal' binary sources.  An important tool
to study the global properties of these populations is their X-ray
Luminosity Function (XLF). \cite{gr03} and \cite{gil04} found that the
XLF of high mass X-ray binaries in the Milky Way, Magellanic Clouds
and nearby starburst galaxies has a smooth, single power-law behaviour
in a broad luminosity range ($10^{36}-10^{40.5}$ erg s$^{-1}$; see
also \citealt{k08}). This suggests that ULXs with luminosity up to a
few $10^{40}$ erg s$^{-1}$ may simply represent the high luminosity
tail of the high mass X-ray binary population.

These pieces of observational evidence, along with the long-term flux
variability and the correlated luminosity/spectral variability
(e.g. \citealt{kub01,b38a,b73}), strongly suggest that a large
fraction of ULXs are accreting BH X-ray binaries with massive
donors. The very high luminosity demands that the accretion rate be
very high, even in case of efficient, disc accretion.
%For ULXs in late type galaxies the accreted
%mass may be supplied by a massive donor and the identification of blue,
%massive stars as the counterparts of some ULXs confirms this
%interpretation.  It is important to stress that a massive donor is not
%the requirement of a specific model, but is needed to fuel persistent
%ULXs irrespectively of the BH mass (e.g. \citealt{pat05,rap05,pz08}).
A massive donor is then needed to fuel persistent ULXs, irrespectively
of the BH mass (e.g. \citealt{pat05,rap05,pz08}), and the
identification of blue, massive stars as the counterparts of some ULXs
confirms this interpretation.  However, transient ULXs associated with
older stellar populations, may be fueled through the rapid accretion
of material accumulated in the accretion disc over a relatively long
period of time, and not necessarily be associated with massive
companions, as it is the case for, e.g, the Galactic BH candidate GRS
1915+105 (\citealt{King02}).

\begin{table*}
 \begin{center}
 \begin{minipage}{140mm}
 \caption{Masses of the BH hosted in some ULXs estimated using X-ray spectroscopic 
 methods.}
 \label{tab1}
 \begin{tabular}{@{}lcccccc}
  \hline
   &  $L_{X,max}^a$ & Edd. limit$^b$ & MCD fit$^c$  & Schwarzschild disc$^d$ & KERRBB fit$^e$ &
   Slim disc fit$^f$  \\
   &  ($10^{40}$ erg s$^{-1}$) & ($M_\odot$) & ($M_\odot$) & ($M_\odot$) & ($M_\odot$) &
   ($M_\odot$) \\
   \hline
% Ho II X-1  & 1.7  & 340 &  &  &  &  \\
 M 81 X-1     & 0.66 & 130 &  &  & $49^{+25}_{-16}$ \\
% M 81 X-9 (Ho IX X-1) & & 220 & $330^{+140}_{-100}$ &  $155^{+60}_{-45}$ &  \\
%              & & 240 & $570^{+270}_{-240}$ &  $270^{+120}_{-120}$ &  \\
 M 81 X-9$^*$  & 1.1 & 220 & $330-470$ &  $150-215$ &  \\
% M 82 X-1     & 17 & 3400 &  &  &  &  \\
 M 101 X-2    & 0.41 & 82 &  &  & $63^{+127}_{-60}$ \\
 NGC 253 X-1  & 0.29 & 58 &  &  & $73^{+22}_{-19}$ \\
 NGC 253 X-3  & 0.1  & 20 &  &  & $63^{+113}_{-31}$ \\
 NGC 1313 X-1 & 2.5 & 500 & $400^{+40}_{-40}$ &  $200^{+40}_{-40}$ &  \\
 NGC 1313 X-2 & 1.5 & 300 & $200^{+160}_{-60}$ & $100^{+80}_{-35}$ &  & 16$\pm$1 \\
% NGC 1313 X-2 & 1.5 & 300 &  & $100^{+80}_{-35}$ &  & 16$\pm$1 \\
 NGC 4559 X-7 & 2.1 & 420 & $2500^{+1950}_{-1100}$ & $1200^{+1000}_{-500}$ &  & 74$\pm$5 \\
 NGC 4559 X-10 & 1.2 & 240 &  &  &  & $31^{+12}_{-9}$ \\
 NGC 5204 X-1 & 0.5 & 100 &  &  &  & 23$\pm$3  \\
% NGC 5408 X-1 & 0.85 & 170 &  &  &  &  \\
  \hline
  \end{tabular}
  \medskip

  $^*$ Also known as Ho IX X-1\par
  $^a$ \cite{b10a,b30a,b44a,RWWGJ05,m07,b50,b36c}.\par
  $^b$ $M_{\rm BH}$ computed from eq.~(\ref{eq1}).\par
  $^c$ $M_{\rm BH}$ computed from eq.~(\ref{eq2}), with $b=9.5$ and $f=1.7$ (\citealt{b44a}).
  The values of $D$ and $K_{BB}$ are taken from \cite{b44a}, \cite{b10a} and \cite{lz08} ($\cos i =1$).\par
  $^d$ $M_{\rm BH}$ computed from eq.~(\ref{eq2}), with $b=19.2$ (Schwarzschild disc; \citealt{lz08})
  and $f=1.7$. The values of $D$ and $K_{BB}$ are from \cite{b44a}, \cite{b10a} and \cite{lz08}
  ($\cos i =1$).\par
  $^e$ $M_{\rm BH}$ from the X-ray spectral fits of \cite{b36c}.\par
  $^f$ $M_{\rm BH}$ from the X-ray spectral fits of \cite{vier06}, multiplied by a 1.3 correction 
  factor (\citealt{vier08}).\par
  \end{minipage}
 \end{center}

\end{table*}

The critical issue is then understanding what is responsible for the
exceptionally high (isotropic) luminosity of these sources.  Two main
scenarios have been proposed.  Firstly ULXs could be relatively normal
stellar-mass BHs ($\la 20 M_{\odot}$) that are either anisotropically
emitting X-ray binaries in a peculiar evolutionary stage
(\citealt{b32,b33}), or are truly emitting above the Eddington limit
via a massive, modified accretion disc structure (e.g.  photon bubble
dominated discs, \citealt{Begel02}; two-phase super-Eddington,
radiatively efficient discs, \citealt{sd06}; slim discs,
\citealt{b15a}), or perhaps via some combination of the two
(\citealt{pout07,King08}).  Secondly, the compact object could simply
be bigger, and the accretion would be in the usual sub-Eddington
regime.  In this case the compact object would be an intermediate mass
black hole (IMBH) with a mass in excess of 100 $M_\odot$
(e.g. \citealt{b11}).  Population synthesis calculations show that, in
both scenarios, the mass transfer rates needed to supply the majority
of the ULXs can be attained over a significant fraction of the life
time of the systems and that the production efficiency of the two
models are comparable (albeit with very large uncertainties on both),
provided that stellar mass BHs can exceed the Eddington limit by a
factor $\ga 10$ (\citealt{pod03,rap05,mad06,mad08}).

%In the following we will shortly review the available estimates of the
%masses of the compact object in ULXs and crititically reconsider the
%stellar-mass versus intermediate-mass black hole interpretation.  We
%suggest that the available data are consistent with a 'third way' in
%which a proportion of ULXs contain $\approx 40-90 M_\odot$ BHs formed
%in a low metallicity environment and accreting in a slightly critical
%regime.
In this Paper we review the available estimates of the masses of the
compact object in ULXs and present a critical re-evaluation of the
current evidence regarding the stellar-mass versus intermediate-mass
black hole interpretation. Here and in a companion investigation
\citep{map09} we highlight an alternative formation scenario, already
suggested before but never explored in depth, in which a proportion of
ULXs contain $\approx 30-90 M_\odot$ BHs formed in a low metallicity
environment and accreting in a slightly critical regime. The plan of
the paper reflects this approach. We start from a quite comprehensive
review of the available estimates of the masses of the compact objects
in ULXs (\S~\ref{sec2}) and critically reconsider the 'traditional'
interpretations of the nature of these sources (\S~\ref{sec3}). We
then discuss the low-metallicity scenario (\S~\ref{sec4}) and some
observational tests to investigate it (\S~\ref{sec4b}). A conclusion
section (\S~\ref{sec5}) follows.

\section{Mass estimates in ULXs: methods and results}
\label{sec2}

Thanks to the identification of the optical counterparts of a handful of ULXs,
%such as Ho II X-1 \citep{b30a}, M 81 X-6 \citep{b40}, or NGC 1313 X-2 \citep{b73}, 
the measurement of the mass function of ULXBs is now well on the way
to becoming possible, and will provide direct constraints on the
masses of individual sources. Unfortunately, given the observational
difficulties associated with such measurements - most notably their
faintness, with typical magnitudes in the range $m_v \sim 22 - 26$
\citep{RLG08}, and the contamination of the counterpart spectrum by
nearby stars - only one very recent measurement of optical periodicity
has been made \citep{liu09}, and a mass function is yet to be
constrained.  However, this is where the observational effort is
focussed at present and where the definitive answer to the question of
whether intermediate or stellar mass BHs power ULXs will come
from. Claims of a radial velocity shift of $\sim 300$ km s$^{-1}$ in
the He {\small II} $\lambda 4686$ line have been reported for the
optical counterpart of NGC 1313 X-2 by \cite{pak06}, but this
measurement may be uncertain (see e.g. \citealt{b49}).

\begin{table*}
 \begin{center}
 \begin{minipage}{140mm}
 \caption{Masses of the BH hosted in some ULXs estimated using timing methods.}
 \label{tab2}
 \begin{center}
 \begin{tabular}{@{}lcccccc}
  \hline
   &  $L_{X,max}^a$ & Edd. limit$^b$ & QPO$^c$ &  Break/Lack of variability$^d$ &  &  \\
    &  ($10^{40}$ erg s$^{-1}$) & ($M_\odot$) & ($M_\odot$) & ($M_\odot$) \\
  \hline
 Ho II X-1  & 1.7 & 340 &  & $\la 100$ \\
 M 82 X-1     & 17  & 3400 & 95--1300 & 25-520 \\
 NGC 5408 X-1 & 0.85 & 170 & 115--1300 & $\sim 100$ \\
 NGC 4559 X-7 & 2.1 & 420 &  & 38-1300 \\
  \hline
  \end{tabular}
  \end{center}
  \medskip

  $^a$ From \cite{b10a,b30a,b49b,b62}.\par
  $^b$ $M_{\rm BH}$ computed from eq.~(\ref{eq1}).\par 
  $^c$ \cite{cas08}.\par
  $^d$ \cite{b10a,dew06,GRRU06,b60}.
  \end{minipage}
  \end{center}
\end{table*}

Until these measurements are performed, we have to rely on indirect
methods to estimate the BH mass.
%Since the identification of ULXs, the
%simplest estimate comes from X-ray luminosity scaling
%arguments. Assuming that the emission of ULXs originates from
%accretion and that it is stationary and isotropic, one might expect
%that the luminosity must be at or below the Eddington limit in order
%for the source to be persistent. 
Assuming that the emission of ULXs originates from accretion and that
it is stationary and isotropic, a lower limit for the BH mass $M_{\rm
BH}$ is obtained for Eddington-limited accretion ($L\approx L_{\rm
Edd}$):
\begin{equation}
\frac{M_{\rm BH}}{M_\odot} \simeq 80 \left( \frac{L}{10^{40} \, {\rm erg \, s}^{-1}} \right)
\simeq 200 \left( \frac{L_X}{10^{40} \, {\rm erg \, s}^{-1}} \right)
\, , \label{eq1}
\end{equation}
where $L_X$ typically refers to the [0.2-10] or [0.3-10] keV band and
we consider a 'bolometric correction' of $\sim 2$ to account for the
flux emitted outside this band. In fact, for the typical spectral
parameters of a bright ULX (colum density $\sim 3\times 10^{21}$
cm$^{-2}$ and power-law photon index $\sim 1.7$) the bolometric flux
is $\approx 2$ times larger than the flux emitted in the 0.2-10 keV
band (e.g. \citealt{pz08}).  This equation is also sensitive to the
material accreted; the Eddington limit will rise by a factor $\sim 2$
for the accretion of helium (e.g. \citealt{gr03}).  Masses of some
ULXs calculated from equation~(\ref{eq1}) are reported in
Table~\ref{tab1}.  Estimates based on this argument may become more
reliable if there is some other evidence that the emission is (almost)
isotropic, as for example when ULXs appear to be responsible for the
photo-ionization of their surrounding optical nebulae
(e.g. \citealt{b30a,abol07}).

Spectroscopic estimates of the BH masses have also been attempted
assuming that the soft component observed in ULX X-ray spectra can be
ascribed to emission from an accretion disc (e.g. \citealt{b44,b44a}).
Assuming that this spectral component can be modelled with the
so-called multicolour disc blackbody model (MCD; \citealt{b48}), it is
possible to express the BH mass $M_{\rm BH}$ as (e.g. \citealt{lz08}):
\begin{eqnarray}
&& \frac{M_{\rm BH}}{M_\odot}=f^2 \frac{67.5}{b} \left(\frac{D}{1 \, {\rm Mpc}}\right) \left(\frac{K_{BB}}{\cos i}\right)^{1/2} \label{A3} \label{eq2}
%&& \frac{\dot{M}}{\dot M_{\rm Edd}}=0.1 f^2 b^2 \left(\frac{D}{1 \, {\rm Mpc}}\right) \left(\frac{K_{BB}}{\cos i}\right)^{1/2} \left(\frac{T_{in}}{\rm 1 keV}\right)^4
\, , 
\label{eq3}
\end{eqnarray}
where $D$ is the distance of the source, $f$ is a colour correction
factor that accounts for transfer effects
(e.g. \citealt{b59,b73a,b14a,b36d}), $b$ is the inner radius in units
of the gravitational radius, and $K_{BB}$ is the MCD normalization
inferred from the spectral fit.
%For a disc terminating at the innermost stable circular orbit,
%$b=6$. Correction factors for $b$ that account for the difference in
%the temperature profile of the MCD with respect to that of the
%standard disc have been derived in the literature (e.g. 
%\citealt{b36,m00,b75}). 
In their work \cite{b44} and \cite{b44a} adopted $b=9.5$ and $f=1.7$
and derived estimates of $M_{\rm BH}$ well in excess of $100 M_\odot$
for M 81 X-9, NGC 1313 X-1 and X-2. A similarly large value of $M_{\rm
BH}$ is obtained also for NGC 4559 X-7 using the MCD spectral
parameters of \cite{b10a} (see Table \ref{tab1}).  However, recently
\cite{lz08} computed appropriate correction factors for $b$ in the
case of a relativistic, standard accretion disc and showed that,
unless the BH is maximally rotating, the BH masses inferred for the
same sources can be significantly lower than the values estimated by
\cite{b44} and \cite{b44a} (see again Table \ref{tab1}).
%In particular, for a Schwarzschild BH, $b=19.2$ and $M_{\rm BH}$ is in the range 
%$\sim 100-200 M_\odot$ (see Table~\ref{tab1}). The only exception is
%NGC 4559 X-7 for which, however, there are hints of timing features in
%the power density spectrum, as we will discuss below.
Smaller BH masses have also been obtained through direct spectral fits of 
relativistic accretion disc models for a sample of disc-dominated ULXs 
by \citet{b36c}.
%We report some of them in Table~\ref{tab1}.

The interpretation of the soft component in terms of emission from a
standard accretion disc suffers from a high degree of equivocality.
%However, apart from the model by \cite{b27}, where it
%arises from the emission, absorption and reflection of a fast ionised
%outflow similar to the warm absorber in AGNs, the other spectral
%models adopted to fit ULX spectra somehow bear the signature of an
%accretion disc, although it can not be used to obtain a direct
%estimate of the BH mass.
Spectral fits with a disc component and a comptonizing corona to the
best available X-ray spectral data indicate that ULXs display distinct
spectral curvature above 2 keV (\citealt{b61,grd09}).  This is because the
corona is optically thick and cool, and hence hides the inner part of
the accretion disc, in what is likely an extreme form of the so-called
very high state of Galactic BH candidates (\citealt{dk06,b61}). In
this case, the soft component is produced by the visible outer regions
of the accretion disc.
%so it cannot be used to estimate the true inner disc radius
%(even taking into account the aforementioned correction factors for $b$). 
Only upper bounds to the inner disc radius can be obtained and,
similarly, the spectroscopic estimates for $M_{\rm BH}$ reported in
Table~\ref{tab1} should be considered as upper limits.  However, as
the thick corona is probably only launched at extreme accretion rates,
this implies that this ``ultraluminous'' state is in the
Super-Eddington regime, and hence the BH masses are relatively small
($< 100 M_{\odot}$; \citealt{b56a}).  Further to this, \citet{grd09}
consider the energy required to launch the thick coronae, and from
that calculate the intrinsic (corona-less) disc temperatures, mainly
recovering temperatures in the correct regime for the discs around
stellar-mass BHs (0.7 - 1 keV).  A subset of ULXs retain apparently
cool discs even after this correction; however, \citet{grd09} argue
this is because these are the highest accretion rate stellar-mass BHs,
in which a strong wind is launched from the central regions, creating
a cool photosphere.  In a similar vein, for other models
characterising super-Eddington accretion, such as the slim disc
(e.g. \citealt{b67,b15a}) or photon bubble models (\citealt{fb07}),
the entire 0.2--10 keV spectrum is produced in the accretion disc,
although its physical state is completely different with respect to
that of a standard disc. However, indirect estimates of $M_{\rm BH}$
can again be obtained from X-ray spectral fits and give values
typically in the range $\approx 15-75 M_\odot$
(\citealt{vier06,vier08}; see again Table~\ref{tab1}).

The mass estimates inferred from X-ray spectral fits of ULXs depend
critically on the interpretation of their spectra which, as mentioned
above, is not unique. The situation will improve in the future as our
understanding of the spectral evolution of ULXs will increase and it
will be possible to select spectral models on the basis of their
consistency with the observed correlation patterns (see
e.g. \citealt{fk09,kp09} for preliminary work on ULX spectral
variability based on {\it XMM-Newton\/} data).  Nonetheless, the
available data points towards BH masses definitely smaller than those
estimated from the early MCD spectral fits.

In the last few years, X-ray timing has provided a new opportunity to
estimate BH masses in ULXs thanks in particular to the detection of
broad band noise and quasi periodic oscillations (QPOs) in the power
density spectrum (PDS) of some ULXs, as M 82 X-1 ($\nu_{QPO}=54-166$
mHz; \citealt{b62a,b49b}) and NGC 5408 X-1 ($\nu_{QPO}\simeq 20$ mHz;
\citealt{b62}).
%The main method adopted is the scaling
%of characteristic frequencies by comparison with those of similar
%timing features observed in Galactic BH candidates. In this respect
%the properties (rms, coherence, noise, variability) of the two QPOs
%observed in M 82 X-1 and NGC 5408 X-1 are strongly reminiscent of the
%so called Type C QPOs in BH candidates (\citealt{b49b,b62}). 
Extrapolating timing and spectral-timing correlations known to exist
for similar timing features in BH binaries and assuming that the
frequency of the QPO scales inversely to $M_{\rm BH}$, various
estimates have been obtained, which are consistent with a rather broad
interval of values: $10-1000 M_\odot$ for M 82 X-1
(\citealt{ft04,b49b,fk07}), $600-5000$ for NGC 5408 X-1
(\citealt{b62}).
%These results are uncertain as presently available data do not allow
%us to establish whether the adopted correlations remain valid over
%such wide ranges of frequencies and luminosities (hence possibly
%masses). However, \cite{cas08} have recently presented a new
%method that overcomes, in part, these difficulties, exploiting the so
%called ``variability plane'', populated by both Galactic black-hole
%candidates and active galactic nuclei. The existence of this
%correlation across several orders of magnitude in mass and luminosity
%means that BH accretion is scale invariant. So, if we make the
%assumption that the accretion flow in ULXs behaves in a similar way to
%other BHs (which remains an open question), using this correlation it
%is possible to derive an estimate of $M_{\rm BH}$ without performing
%critical extrapolations outside their range of validity, although it 
%is still necessary to extend to the ULXs a relation between 
%characteristic variability frequencies known to hold only for 
%stellar-mass compact objects.
Recently, a new timing approach has been presented to assess BH masses
in ULXs, which is based the so called ``variability plane'', populated
by both Galactic black-hole candidates and active galactic nuclei.
Assuming that the accretion flow in ULXs behaves in a similar way
(which remains an open question) and taking into account the
uncertainty on the efficiency of the accretion disc, \cite{cas08} find
that $M_{\rm BH}$ is in the interval $\sim 95-1300 M_\odot$ for M 82
X-1 and $\sim 115-1300 M_\odot$ for NGC 5408 X-1 (see
Table~\ref{tab2}). In combination with QPOs, the scaling of the break
frequencies of the broad band noise, by comparison with the
corresponding timing features of Galactic BH candidates, has also been
proposed for estimating $M_{\rm BH}$. This method has been applied to
NGC 5408 X-1 and gives a similar range of masses to QPOs (see
Table~\ref{tab2}). A tentative identification of a break at $\sim 28$
mHz in the PDS of NGC 4559 X-7 has also been reported
(\citealt{b10a}), although \cite{bar07} called it into question.  The
inferred BH mass is reported in Table~\ref{tab2}. 

The problem with using timing properties of ULXs to directly infer
masses is that it remains to date unclear how exactly these quantities
are related. All the estimates are based upon tentative
identifications of the timing features and the use of scaling laws
that are known to hold only for a limited number of objects, and are
therefore highly uncertain.

A further possibility is to use the non-detection of variability power
in the PDS to limit the size of the BH, assuming that all power is at
higher frequencies, as is seen in various XRB states. Adopting this
approach \cite{GRRU06} found $M_{\rm BH} \la 100 M_\odot$ for Holmberg
II X-1 (see again Table~\ref{tab2}).  In fact, \citet{hvr09} have
demonstrated that suppressed temporal variability (compared to the PDS
of classic XRBs and AGNs) appears a common feature of ULXs.  There are
several possible explanations of this - the variability is limited to
higher frequencies, the data in the {\it XMM-Newton\/} band pass are
disc-dominated (and so any variability is heavily diluted) or the ULXs
are in a new, super-Eddington accretion state in which the X-ray
emission is very stable (note this is predicted in the hydrodynamic
simulations of highly super-Eddington accretion by \citealt{ohsuga07}).
Again, the common thread running through all these models is that the
BH is relatively small.

\section{Intermediate or stellar-mass BHs?}
\label{sec3}

%Assuming that ULXs are BH binary systems, the very high luminosity
%demands that the accretion rate be very high, even in case of
%efficient, disc accretion. For ULXs in late type galaxies the accreted
%mass may be supplied by a massive donor and the identification of blue,
%massive stars as the counterparts of some ULXs confirms this
%interpretation.  It is important to stress that a massive donor is not
%the requirement of a specific model, but is needed to fuel persistent
%ULXs irrespectively of the BH mass (e.g. \citealt{pat05,rap05,pz08}).
%A massive donor is then needed to to fuel persistent ULXs, irrespectively 
%of the BH mass (e.g. \citealt{pat05,rap05,pz08}).
%However, transient ULXs associated with older stellar populations, may
%be fueled through the rapid accretion of material accumulated in the
%accretion disc over a relatively long period of time, and not
%necessarily be associated with massive companions, as it is the case
%for, e.g, the Galactic BH candidate GRS 1915+105.

We have reviewed the mass estimates drawn from observations of
individual ULXs; we now ask how these results fit into our more
general understanding of the possible nature of their underlying
engines.  ULX models differ mainly in the assumptions on the physical
state of the disc and its mode of accretion. If the accretion disc is
in a standard regime, then emission is isotropic and the most
straightforward interpretation of the exceptionally high luminosity of
ULXs is that they contain BHs with large masses. Both the very high
luminosity and low characteristic temperature (and high normalization,
see previous Section) of the soft spectral component have been taken
as evidence for this interpretation.
%As discussed above mass estimates in excess of $100 M_\odot$ are
%derived from the luminosity and the characteristic
%temperature/normalization of the soft X-ray spectral component,
%assuming that it represents emission from an accretion disc.

But how big is the BH mass? Early estimates based on
equations~(\ref{eq1}) and~(\ref{eq2}) gave masses largely in excess of
$100 M_\odot$ (up to several thousands; see Table~\ref{tab1}). The
obvious question is then how a BH this massive may form. It has been
proposed that remnants from the collapse of Population III stars
formed in cosmological epochs (\citealt{mr01}) may trigger ULX
activity, if they can capture a donor star to accrete from.  A second
formation route for IMBHs may be in globular clusters, through
repeated mergers of stellar mass BHs \citep{mh02}, or in young, dense
stellar super clusters, from the dynamical collapse of supermassive
stars in their centres (e.g. \citealt{portzw04}). ULX activity would
be sustained by binary companions captured in the cluster
\citep{bl06}. The latter may be a possible explanation for some ULXs
(e.g. M82 X-1) but, in galaxies with starburst activity, the majority
do not appear inside such supermassive clusters
(e.g. \citealt{zez02,k04}).  Another difficulty with the IMBH
interpretation is the apparent break in the power-law slope of the XLF
of the high mass X-ray binary population in external galaxies at a
luminosity $\sim 2\times 10^{40}$ erg s$^{-1}$.  If there is a
population of large IMBHs contributing to the XLF, the break suggests
that they turn off at this luminosity; yet this is at a rather low
fraction of the Eddington limit for putative large IMBHs ($\sim 10\%$
for a $1000 M_{\odot}$ BH).  This would be rather unusual behaviour,
as no other accreting class switches off at only a fraction of their
Eddington limit (\citealt{b56a}). Finally, the co-location of ULXs
with regions of star formation, such as those in the Antennae and
Cartwheel galaxies, implies that they must be (relatively)
short-lived, which requires successive generations of ULXs to be
formed over the duration of the star formation event. Thus, if all
ULXs in these regions were IMBHs, an unfeasibly large fraction of star
forming mass would end up in IMBHs (\citealt{King04,map08}).  In
principle, these arguments rule out all but a small minority of ULXs
from being IMBHs bigger than $\sim 100 M_{\odot}$.

So can we explain ULXs as stellar-mass BHs? If the accretion flow in
ULXs is in a different regime, the situation may be different and
either the isotropy and/or the Eddington limit may be
circumvented. This occurs if the accretion rate is at or above
Eddington, so that radial advection of thermal and radiative energy
(slim disc; \citealt{b1,b15a}) and/or radiation-driven instabilities
(photon bubble model; \citealt{Begel02,Begel06}) set in. A slim disc
can sustain larger accretion rates than a standard disc and modest
super-Eddington luminosities ($\la 10 L_{\rm Edd}$). As the emission
is isotropic, luminosity scaling arguments similar to those discussed
above give $M_{\rm BH} \ga 20 (L_X/10^{40} \, {\rm erg \, s}^{-1})
M_\odot$. Therefore, the presence of slim discs would imply that only
the brightest known ULXs (with $L_X \ga 3\times 10^{40}$ erg s$^{-1}$)
need BHs significantly more massive than the stellar-mass BHs in our
Galaxy (see also Table~\ref{tab1}). Accretion discs dominated by
photon bubble transport may also reach super-Eddington luminosities,
while remaining geometrically thin. For a $20 M_\odot$ BH, the maximum
luminosity is $\sim 30 L_{\rm Edd}$ (\citealt{Begel06}).  This may in
principle account for the emission of all but the very brightest ($\ga
5\times 10^{40}$ erg s$^{-1}$) ULXs, but photon bubble-dominated discs
are subject to the same thermal and viscous instabilities that
characterize the inner region of radiation pressure dominated discs
and hence may be significantly unsteady (with more than a factor of 10
variation in the emitted flux) on short time scales
(e.g. \citealt{b73a}). However, bright ($\ga 10^{40}$ erg s$^{-1}$)
ULXs typically show more limited fluctuations in the observed X-ray
luminosity.  An alternative scenario for generating steady,
super-Eddington luminosities from stellar-mass BHs, is the radiatively
efficient, two-phase super-Eddington accretion disc model by
\cite{sd06}. In this model the gravitational potential energy is not
trapped in the disc, but effectively removed from it through magnetic
buoyancy and dissipated in a corona.
%This would lead to a steady behaviour. Confining magnetic 
Fields anchored in the disc and/or Compton drag in the low density
corona prevent the launching of a wind, keeping the radiative
efficiency high and assuring super-Eddington luminosities. However, we
note that there are many assumptions and theoretical uncertainties
that need to be clarified in this model, and a quantitative estimate
of the maximum attainable luminosity is not yet available.

If accretion becomes largely super-Eddington, other processes may be
present that complicate the picture. A thick disc may form and the
emission becomes beamed (\citealt{b32,King02}). At the same time,
outflows and powerful ejection of matter along the axis perpendicular
to the accretion disc may be produced (as in SS433;
e.g. \citealt{pout07}).
%These super-Eddington rates
%may occur in particular phases of the evolution of a binary system,
%for example when the donor has a radiative envelope and is more massive than the
%BH (thermal timescale mass transfer; \citealt{b32}).
In these assumptions, emission is no longer isotropic. If $L_{iso}$ is
the apparent isotropic luminosity, the BH mass inferred from
equation~(\ref{eq1}) must then be corrected for the so called beaming
factor $b_f$, that represents the fractional opening angle of the
beam: $M_{\rm BH} \simeq 20 (b_f/0.1) (L_{0.2-10,iso}/10^{40} {\rm
erg} \, {\rm s}^{-1}) M_\odot$. Hence, simple beaming can not account
for bright ($\ga 2 \times 10^{40}$ erg s$^{-1}$) ULXs, unless one is
willing to consider a beaming factor $< 0.05$ and therefore a very
narrow opening angle ($< 18^{\circ}$), that appears to be more
consistent with a jet rather than a geometrical funnel in a thick
disc. However, in addition to having beamed emission, thick discs may
also radiate at super-Eddington luminosities, reaching at most
$(1+ln({\dot{M}}/{\dot M_{\rm Edd}})) L_{\rm Edd}$
(e.g. \citealt{pout07,King08}). A hypothetical ULX in this state would
bear some similarity to a Galactic BH candidate in the very high
state, albeit with the ULX being in a much more extreme version of
this state, with powerful winds carrying away the excess matter and
energy, potentially thickening the corona and even producing a cool
photosphere (consistent with the X-ray spectral modelling of
\citealt{grd09}).
%similar to GRS 1915+105 and XTE J1550-564 (although they do not remain in this %state for very long).
%Even with a combination of a reasonable beaming factor 
%$b_f\simeq 0.5$ and super-Eddington emission it is hard to 
%explain ULXs with luminosities $\ga 10^{40}$ erg s$^{-1}$ assuming 
%accretion onto a stellar-mass BH. In order to account for the high
%luminosity tail of the ULX population, recently \cite{King09} proposed 
%that the beaming factor may depend on the accretion rate as $b_f 
%\propto ({\dot{M}}/{\dot M_{\rm Edd}})^{-2}$.
A combination of a beaming factor $b_f\simeq 0.3-0.5$ and super-Eddington 
emission ($\sim 10 L_{Edd}$) may in principle explain ULXs with 
luminosities up to $\approx 10^{40}$ erg s$^{-1}$, assuming accretion onto a 
stellar-mass BH. Furthermore, if a dependence of the beaming factor on the 
accretion rate is assumed ($b_f \propto ({\dot{M}}/{\dot M_{\rm Edd}})^{-2}$;
\citealt{King09}), it might be possible to account also for the high
luminosity tail of the ULX population. 
%This meas that the apparent luminosity of a 
%ULX goes up rapidly as ${\dot M}$ increases.
%However, the break in the XLF of ULXs at $\sim 2\times 10^{40}$ erg s$^{-1}$ 
%appears to be at odd with this open-ended beaming model, as it implies a 
%hitherto unexplained cut-off at $\sim 20-30 {\dot M_{\rm Edd}}$ in the 
%maximum achievable accretion rate.

%\cite{rap05} and \cite{mad06} studied the
%production efficiency of ULX sbinary systems with a massive donor and
%a stellar mass BH and found that super-Eddington mass transfer rates
%can be sustained over a significant fraction of the lifetime of the
%system. However, it is unclear if such systems can indeed account for
%the observed number of ULXs once orientation effects caused by beaming
%are taken into account. 
%This is even more true for hypercritically accreting ULXs. 
%Another difficulty with beamed models is the fact that
%unreasonably small beaming factors ($\la 3\times 10^{-2}$) are
%required to explain ULXs with $L_{2-10} \ga 3\times 10^{40}$ erg
%s$^{-1}$, which are difficult to achieve in the simple geometrical
%funneling scenario of Super-Eddington accretion. 
%Significant beaming is also ruled out by the detection of QPOs in M 82 X-1 
%and NGC 5408 X-1, as their are a coherent signal.
%Finally, some ULXs are embedded in very extended, energetic nebulae
%that are photo-ionised by the radiation emitted by the ULX. In the
%case of Ho II X-1, the He II flux implies a lower bound on the X-ray
%luminosity in the range $4-6 \times 10^{39}$ erg s$^{-1}$ and hence a
%BH mass in excess of $\sim 20-40 M_\odot$ (\citealt{b30a}).

In order to assess the viability of the different scenarios for the
origin of ULXs, it is necessary to understand the possible
evolutionary history of the various types of candidate binary systems
and compare them with the available observations.  Calculations of the
evolutionary tracks of ULX binaries and model population studies of
systems containing stellar-mass BHs and IMBHs have been carried out by
several authors (e.g. \citealt{pod03,pat05,rap05,mad06}).
%obtaining statistical estimates of the expected number of ULXs per galaxy 
%and of their production efficiency.
%Statistical estimates of the expected number of ULXs per galaxy 
%and of their production efficiency have been derived for systems containing 
%both stellar-mass and IMBHs. 
Although calculations depend sensitively on uncertain parameters of
the common envelope phase, it turns out that stellar-mass BHs
accreting at super-Eddington rates may be able to account for most of
the observed ULXs (except for the brightest), if they violate the
Eddington limit by a factor $\sim 10-30$ (\citealt{pod03,rap05}).
%Similarly, IMBH systems with a representative
%mass of 1000 $M_\odot$ have an acceptable production efficiency of bright
%($L_X \ga 10^{40}$ erg s$^{-1}$) ULXs if the donor star is above 
%$\sim 8 M_\odot$ and the initial orbital separation is small ($\la
%6-40$ times the initial main sequence radius of the donor; \citealt{mad06}).
Similarly, IMBH systems might produce bright ($L_X \ga 10^{40}$ erg
s$^{-1}$), persistent ULXs and have an acceptable production
efficiency if the donor star is $\ga 10 M_\odot$ and the initial
orbital separation is small ($\la 6-40$ times the initial main
sequence radius of the donor; \citealt{pat05,mad06}).
%However, their expected number is rather uncertain, as it depends on the
%unknown formation mechanism and the probability of capturing a companion 
%into a close orbit.

%For IMBHs with mass $\ga 100 M_\odot$ the average mass transfer rate needed 
%to sustain persistent ULX emission at the observed luminosities is
%$\sim 10-15 M_\odot$ (\citealt{pat05}).
%Therefore, population studies suggest 
%that, under certain assumptions on binary synthesis calculations and
%IMBH formation, both scenarios may be consistent with the observed number of 
%ULXs, except for the very most luminous ones for which extremely large violations of 
%the Eddington limit are required within the framework of stellar-mass BH models.

Along the same lines, theoretical calculations of the color-magnitude
(CM) diagrams for systems containing stellar-mass BHs and IMBHs are
used to constrain ULX models and their evolutionary history by
comparison with observations of their optical counterparts
(\citealt{mad08,pz08}). As already mentioned, in order to supply the
mass transfer rates needed to fuel ULXs, rather massive donor stars
are required.  The evolutionary tracks of such systems are strongly
affected by the binary interaction and the emission from the accretion
disk, including X-ray irradiation. Numerical computations show that
the regions of the CM diagram with the highest probability of finding
ULX optical counterparts have $B-V$ between $\sim -0.1$ and $-0.3$ and
correspond to the early phases of the evolution of massive donors,
while they are on the main sequence or the subgiant branch
(\citealt{mad08}). This result is consistent with the properties of
the observed counterparts. Similarly, the most favourable orbital
periods are between 1 and 10 days, corresponding again to the main
sequence or subgiant phases. This is true for both stellar-mass BH and
IMBH systems and depends on the fact that the donors spend most of
their life time in these phases.
%Therefore, normal nuclear-driven mass transfer is sufficient
%to provide high enough mass transfer rates to sustain the ULX emission, although
%in some rare cases the evolution may be driven by the thermal timescale mass 
%transfer.
In these conditions normal nuclear-driven mass transfer is effective and 
provides sufficiently high transfer rates to sustain the ULX emission, 
although in rare circumstances the evolution may be driven by the thermal 
timescale mass transfer during the giant phase.
%Comparison of theoretical CM diagrams with the colors and magnitudes
%of observed optical counterparts shows that IMBH systems are somewhat favoured
%with respect to stellar-mass BH binaries, as they are more luminous (\citealt{mad08,pz08}). 
%This is because they may contain more 
%massive donors ($\ga 25 M_\odot$) and have more extended accretion discs, that dominate 
%the optical emission. On the other hand, stellar-mass BH systems with stable
%mass transfer contain, on average, less massive donors and have lower luminosities.

\begin{figure*}
 \includegraphics[width=14cm]{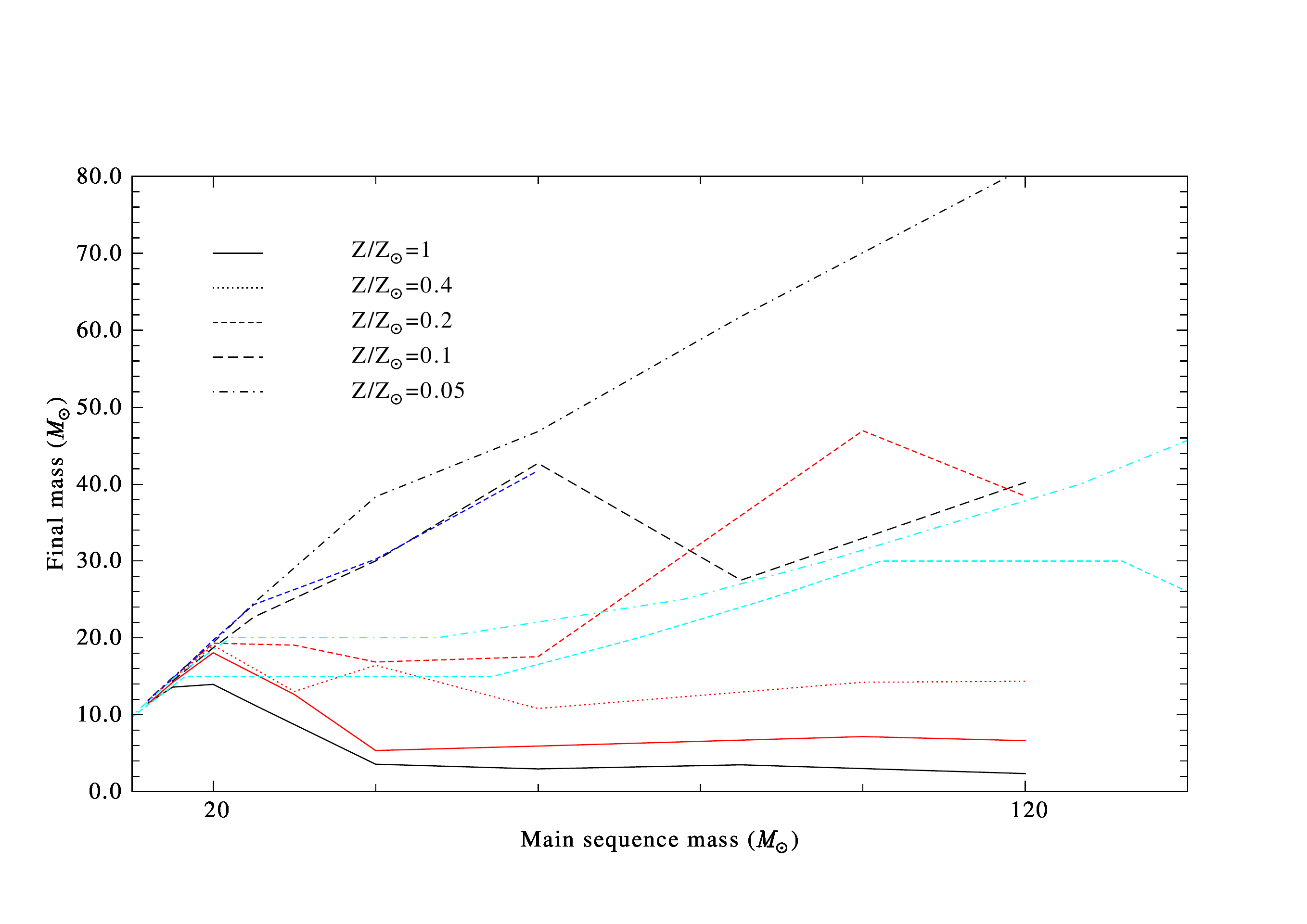}
 \caption{
%Mass at the terminal age main sequence (TAMS; {\it red}) as a 
%function of initial mass for stars at different metallicities. 
%Also shown is the mass lost in line-driven winds ({\it blue}). 
%Mass loss rates are taken from
%Nieuwenhuijzen \& de Jager (1990), with a $\propto Z^{0.5}$ scaling 
%law for the metallicity dependence (Nugis \& Lamers 2000).
Final mass as a function of initial mass for stars of different metallicities 
as computed by Maeder (1990, 1992; {\it black}), Maeder \& Myenet (2001; 
{\it blue}), Portinari et al. (1998, {\it red}), and Eldridge \& Tout (2004,
{\it cyan}).
}
 \label{fig2}
\end{figure*}

%%%%%%%%%%%%%%%%%%%%%%%%%%%%%%%%%%%%%%%%%%%%%%%%%%%%%%%%%%%%%%%%%%%%%%%%%%%%%
%
%
%
%
\section{A different interpretation}
\label{sec4}

%Soon after their identification, the possibility that, because of
%their unusually high luminosity, ULXs could host IMBHs
%of $10^2-10^4 M_\odot$ was welcomed with great enthusiasm by the
%scientific community as they could represent the massive BHs that were
%believed to form through dynamical interactions in dense, massive
%stellar clusters and could also be excellent sources of gravitational
%radiation (\citealt{mh02}). The subsequent proposal that they
%could actually be beamed X-ray binaries with compact objects of mass
%comparable to that of Galactic BHs ($7-20 M_\odot$) has led to a
%polarisation of the discussion around these two opposite scenarios.

A critical revaluation of the available observational evidence
presented in Sections~\ref{sec2} and~\ref{sec3} indicates that
%despite the ULX production efficiency of IMBH binaries may be comparable 
%to that of stellar-mass BH ones,
BHs of several hundreds to 
thousands $M_\odot$ are not required for the majority of ULXs.
However, it is not possible to rule out that they are present in the
handful of known hyper-luminous ($\sim 10^{41}$ erg s$^{-1}$) objects
and/or in the sources showing large amplitude broad band noise or QPOs
in their PDS (such as M 82 X-1 and NGC 5408 X-1; see Table~\ref{tab2}).
At the same time, models with stellar mass BHs may work for a large
fraction of the ULX population, if the accretion flow has some degreee
of beaming and is super-Eddington. Bright ($\ga 10^{40}$ erg s$^{-1}$)
ULXs may be accounted for if some form of modified beaming that
accounts for a suitable dependence of $b_f$ on the parameters of the
accretion flow is allowed \citep{King09}. Although none of these
scenarios can be ruled out, the fact that the observational limits
discussed in Section~\ref{sec2} are converging towards masses $\la 100
M_{\odot}$, but bigger than stellar mass BHs, led us to consider an
alternative interpretation. In our scenario bright ULXs may contain
BHs with masses above 30-40 $M_\odot$ and up to $\sim 80-90 M_\odot$,
formed from ordinary stellar evolution of massive ($30-120 M_\odot$)
stars in a low metallicity natal environment. While this idea has
already been suggested before (e.g. \citealt{b55,b10a,b73}), it has
not yet been explored quantitatively in detail.

Stars with mass $\ga 8 M_\odot$ produce compact remnants from the
gravitational collapse of the iron core. For stars up to $\sim 25-30
M_\odot$ the collapse is halted when the core reaches nuclear
densities: the star explodes and a neutron star forms. For larger main
sequence masses, the early acccretion of the inner mantle onto the
core before shock passage and the fallback of material afterwards
cause the newly formed proto-neutron star to collapse to a BH after
the supernova explosion (\citealt{zamp02} and references therein). At
solar metallicity, these fallback BHs reach at most $\sim 10 M_\odot$
as, for very massive stars with mass $\ga 40 M_\odot$, the stellar
envelope is in large part effectively removed through line-driven
winds, while the remaining part is expelled by the supernova
explosion.

For sub-solar metallicities, however, this mechanism becomes
progressively less efficient and stars with masses above $\sim 30-40
M_\odot$ may retain rather massive envelopes at the time of
explosion. The supernova shock wave then loses more and more energy in
trying to unbind the envelope until it stalls and most of the star
collapses to form a BH with a mass comparable to that of the
pre-supernova star \citep{fr99}. These may be the BHs hosted in some
ULXs. Their mass would not be significantly larger than $\sim 80-90
M_\odot$ as above $\sim 100-120 M_\odot$ a star undergoes pulsational
pair-instability in its core and most of the envelope mass is
expelled. We note that the possibility of forming BHs in this mass
range through a different channel (binary mergers of massive
components) was also proposed few years ago (e.g. \citealt{bsr04}).

Computation of the evolution of massive stars up to advanced
evolutionary stages for different metallicities and/or including mass 
loss have been performed by several authors
(e.g. \citealt{hv81,mae90,mae92,por98,heger03,chief04,eld04,h07}).
%The mass loss history of a star plays a major role in determining its
%final evolution. 
These works adopted known empirical parameterizations of the mass loss
rate for stars over the whole Hertzsprung-Russell diagram
(e.g. \citealt{dejag88,ndj90}).
%that are still considered very accurate \citep{cro01}
A certain degree of uncertainty is introduced if the star enters some
peculiar evolutionary stages, such as the Wolf-Rayet (WR) stage
(e.g. \citealt{lan89,wel99}). In fact, mass loss rates in WR and O
stars may be significantly reduced if the wind is clumpy as a
consequence of, e.g., supersonic turbulence \citep{mc94,ful06}.
%A decrease by a factor of $\ga 3$ in 
%mass loss with respect to homogeneous wind models is attainable during the WR phase. 
%Clumpy winds in WR stars may lead to a decrease by a factor of $\ga 3$ in 
%mass loss with respect to homogeneous wind models.
In particular, during the WR phase a decrease by a factor of $\ga 3$
with respect to homogeneous wind models is attainable.  This turns out
to be compatible with some observational estimates and would clearly
lead to more massive pre-supernova stars and hence more massive
BHs. Additional uncertainty is caused by the dependence of mass loss
on metallicity. A scaling law $\propto Z^{0.5}$ is often adopted for
hot stars (see e.g. \citealt{kud89,nl00}). For example, at the end of
main sequence, the mass of a star with an initial mass of $100
M_\odot$ is $\sim 25 M_\odot$ at solar metallicity and $\sim 75
M_\odot$ for $Z\approx 0.1 Z_\odot$.  However, the mass lost during
the He burning phase may be much more significant than that occurring
during main sequence. According to the adopted mass loss history, the
final mass of a star (after all nuclear burning stages are exhausted)
may differ up to a factor of $\sim 2$ (or even more for clumpy winds)
for a given metallicity.  Considering again a star with an initial
mass of $100 M_\odot$, Figure~\ref{fig2} shows that its final mass may
be in the interval $\sim 3-6 M_\odot$ for $Z\approx Z_\odot$ and $\sim
30-70 M_\odot$ for $Z\approx 0.1 Z_\odot$.

As already mentioned above, the final evolutionary stages of a star
and, in particular, the outcome of the final collapse depend
critically on how massive is the envelope that it retains at the time
of explosion. Therefore, owing to their larger final masses, the fate
of stars with sub-solar metallicity is likely to be be quite different
from that of higher metallicity stars. Although different authors
obtain different results for the mass of the compact remnant, it is
not unreasonable to think that, if an envelope more massive than $\sim
30-40 M_\odot$ is retained at the time of explosion, a low metallicity
($Z\approx 0.1 Z_\odot$) star may collapse directly to form a BH of
comparable mass. Looking again at Figure~\ref{fig2}, it is possible to
see that, already at metallicities $\la 0.2 Z_\odot$, stars with
initial mass $\ga 60 M_\odot$ appear to possess final envelope masses
that overcome this threshold for direct BH formation.  Significant
stellar rotation (hundreds of km s$^{-1}$) may change this picture
somewhat, as it enhances the mixing of heavy elements throughout the
star increasing the metal content of the envelope and consequently
mass loss (e.g. \citealt{mm01,mm05}). Also, ejection of part of the
envelope during the final collapse can not be ruled out. However, if
the core is not rapidly rotating, there is no good reason why most of
the star should not collapse into a BH. Therefore, the formation of a
$\ga 30-40 M_\odot$ BH throughout the evolution of a low metallicity,
slowly rotating star of $40-120 M_\odot$ appears a viable 
possibility. The parameter space in the metallicity-main sequence mass plane where
this formation channel may actually work corresponds to the
black-shaded area between $\sim 40$ and $\sim 100 M_\odot$ in Figure~1
of \cite{heger03} (see also Figure~5 in \citealt{eld04})\footnote{An estimate of the remnant mass 
from massive stars and its dependence on metallicity and wind mass loss 
rates has been very recently derived also by \cite{b09}.}.

At variance with intermediate mass BHs, the formation of these very
massive stellar remnant BHs does not require an exotic, new mechanism
but is referable to ordinary stellar evolution. Also, the unbroken
power-law slope of the X-ray binary population up to $\sim 2 \times
10^{40}$ erg s$^{-1}$ is consistent with the fact that variations in
metallicity may produce a continuum distribution of BH masses from the
stellar-mass BHs of $10-20 M_\odot$ up to the suggested BHs of
$\approx 40-90 M_\odot$. Given the size of these BHs, no difficulty
with the fraction of star-forming mass in large starbursts ending up
in BHs would arise, and so the objection of \cite{King04} for ULXs as
$\sim 1000 M_{\odot}$ IMBHs is circumvented. At the same time, only
modest beaming ($b_f\sim 0.5$) or slight violations of the Eddington
limit (a factor of a few) would be needed to account for the
luminosity of bright ($\ga 10^{40}$ erg s$^{-1}$) ULXs, at variance
with the extreme accretion scenarios required by stellar mass BH
models. Also the essentially isotropic irradiation of X-ray
photoionised nebulae would find an explanation.

\section{Observational tests}
\label{sec4b}

Model population studies for our scenario are needed to determine the
production efficiency of binary systems containing very massive BHs
and to understand if they are in agreement, in a statistical sense,
with the available X-ray and optical data of ULXs. However, there may
be already some indications strengthening our present suggestion. In a
parallel, preliminary investigation, we show that massive BHs formed
in low metallicity environments might well explain most of the ULXs
observed in the Cartwheel galaxy \citep{map09}. Also, the optical
luminosities of massive BH systems would be, on average, larger than
that of stellar-mass BHs, as the former allow for more massive donors
($\ga 25 M_\odot$) and have more extended accretion discs that
dominate the optical emission. This would make them more consistent
than stellar-mass BHs with the observed distribution of the luminosity
of the ULX optical counterparts (e.g. \citealt{mad08,pz08}).

A crucial aspect of the interpretation of ULXs in terms of BHs from
the direct collapse of low-$Z$, massive stars is the metallicity of
the environment in which ULXBs form. The available estimates appear to
favour a low metallicity scenario, although there are some
discrepancies. Optical observations appear to provide evidence of
sub-solar metallicity in the environment of some ULXs. The emission
nebula surrounding Ho II X-1 has a spectrum resembling that of a
high-excitation H\,{\small II} region typical for low metallicity
($Z\sim 0.1 Z_\odot$) star-forming regions (\citealt{mir02,b55}).  The
stellar environment of NGC 4559 X-7 shows a blue-to-red supergiant
ratio ($3\pm 1$) and colours of the red supergiant population
consistent with a low metal abundance environment with $Z=0.1-0.4
Z_\odot$, similar to that in the SMC and other nearby dwarf galaxies
(\citealt{sor05}). Also the stellar field around NGC 1313 X-2 is
characterised by a low metallicity ($[Fe/H]=-1.9\pm 0.3$), as inferred
from the intrinsic colour of the red giant branch (\citealt{grise08});
studies of the metal abundance of H\,{\small II} regions in NGC 1313
also give low values of Z ($\sim 0.008$; \citealt{wr97,hc07}).

\cite{b72} analysed high signal-to-noise {\it XMM} spectra
of a sample of 14 ULXs, trying to determine the Oxygen abundance from
the detection of K-shell photoionization edges. They apparently find
values that match the solar abundance. However, the comparison of the
X-ray estimates with a compilation of $[O/H]$ ratios determined
through spectrophotometric studies of H\,{\small II} regions
(\citealt{pil04}) and with the luminosity-metallicity relation derived
from the Sloan Digital Sky Survey (\citealt{trem04}) shows significant
systematic differences, especially at low galaxy luminosities and
sub-solar metallicities.  We note that Ho II X-1 provides a clear
example of this dichotomy for a ULX; its {\it XMM-Newton\/} RGS
spectrum suggests a significantly higher metallicity ($\sim 0.6$ times
solar; \citealt{GRRU06}) than the optical data.  At the same time,
X-ray spectral fits in at least the case of NGC 4559 X-7 seem to
provide evidence for a more subsolar metallicity ($Z\sim 0.3 Z_\odot$;
\citealt{b10a}).

It will be possible also to test our proposal against the stellar-mass
BH interpretation as it will lead to a different spatial distribution
of bright ULXs. In fact, ULXs from stellar-mass BHs should essentially
appear anywhere in regions of star formation or in young stellar
environments, regardless of metallicity.  Indeed there may even be a
bias towards these objects appearing in low metallicity regions, as
effects such as the low mass loss rate in stellar winds will tend to
keep binaries close, allowing more high mass transfer systems.
However, if we make the reasonable assumption that the modes of
accretion and Eddington rates will be similar in both standard $\sim
10 - 20 M_{\odot}$ black holes, and the $\sim 30 - 90 M_{\odot}$ black
holes examined in this paper, then the latter black holes may simply
be on average {\it brighter\/}.  Hence, in our proposed scenario ULXs
should show some evidence of correlation (in terms of position and
average luminosity) with low metallicity environments. So, one of the
definitive tests of our proposal would be to survey ULX locations, and
determine whether a relationship between ULX luminosity and local
metallicity was evident in a large enough sample to provide
statistically meaningful results.

We note that evidence supporting our argument is already available.
Firstly, \cite{sst08} have recently surveyed galaxies within the Local
Volume and determined that the specific ULX frequency decreases with
host galaxy mass above $\sim 10^{8.5} M_{\odot}$.  This means that
smaller, lower metallicity systems have more ULXs per unit mass than
larger galaxies, consistent with the idea that BHs can at least form
and/or feed more efficiently in low metallicity environments.
Secondly, we note that there is interesting evidence from our own
relative backyard that the brighter ULXs are more abundant in the
late-type spiral galaxies we might expect to be low-metallicity
systems.  The {\it ROSAT\/} High Resolution Imager ULX survey of
\cite{b41} lists 15 ULXs within 5 Mpc\footnote{Their survey covers 27
galaxies within 5 Mpc, 17 of which are of Hubble type Sd or later.},
four of which have observed X-ray luminosities in excess of $5 \times
10^{39}$ erg s$^{-1}$.  All four of the very luminous ULXs reside in
galaxies of Hubble type Sd or later.  This compares to only 3 of the
11 lower-luminosity ULXs being hosted by similarly late systems - the
remainder are in galaxies of type between Sab - Scd.  Although the
sample of ULXs is small, this clearly supports the case that brighter
ULXs preferentially occur in the smaller, lower-metallicity systems
where we might expect to find the very massive stellar remnant BHs.

It is worth noting that, recently, \cite{prest07} and \cite{sf08}
succeeded in performing dynamical mass measurements using Gemini and
Keck spectra of the Wolf-Rayet optical counterpart of IC 10 X-1, a
variable X-ray source in the the Local Group metal poor starbust
galaxy IC 10. They find a BH mass in the range $23-33 M_\odot$, which
represents the most massive BH known to exist in a binary system and
definitely corroborates our interpretation.

%%%%%%%%%%%%%%%%%%%%%%%%%%%%%%%%%%%%%%%%%%%%%%%%%%%%%%%%%%%%%%%%%%%%%%%%%%%%%
%
%
%
%
\section{Conclusions}
\label{sec5}

In the last few years, X-ray and optical observations have
significantly boosted our understanding of ULXs. We are now confident
that the majority of these sources are X-ray binaries in external
galaxies and we suspect that many may have massive binary
companions. Yet, the most fundamental questions on ULXs still remain
to be definitively answered: do they contain stellar or intermediate
mass BHs? How do they form?

A critical revaluation of the available evidence presented indicates
that BHs of several hundreds to thousands $M_\odot$ are not required
for the majority of ULXs, although they might be present in the
handful of known hyper-luminous ($\sim 10^{41}$ erg s$^{-1}$) objects
and/or some sources showing timing features in their power density
spectra. At the same time, however, stellar mass BHs may be quite a
reasonable explanation for ULXs below $\sim 10^{40}$ erg s$^{-1}$, but
they need super-Eddington accretion and some suitable dependence of
the beaming factor on the accretion rate in order to account for ULXs
above this (isotropic) luminosity.
%The observational limits
%from current data, particularly those coming from X-ray spectroscopy,
%are consistent with bigger BHs.

We investigated in detail an alternative scenario in which bright ULXs
contain BHs with masses above $\sim 30-40 M_\odot$ and up to $\sim
80-90 M_\odot$, produced by stars with initial, main sequence mass
above $\sim 40-50 M_\odot$. At sub-solar metallicity, the explosion
energy of these stars is not sufficient to unbind the envelope and
most of the star collapses to form a BH with a mass comparable to that
of the pre-supernova star. These may be the BHs hosted in bright
ULXs. Above $\sim 100-120 M_\odot$ pulsational instability becomes
effective and most of the envelope is expelled from the star.

The formation of these very massive stellar remnant BHs does not
require an exotic, new mechanism but is referable to ordinary stellar
evolution. For luminosities $\sim 10^{40}$ erg s$^{-1}$, this would
imply only modest violations of the Eddington limit, attainable
through very modest beaming ($b_f\sim 0.5$) and/or slightly
super-critical accretion.
%This scenario would overcome many of the
%problems encountered by stellar-mass and intermediate-mass black hole
%interpretations.

Measurements of the metallicity of the environment of some ULXs appear
to favour a low metallicity scenario, although there are some
discrepancies. Surveys of ULX locations looking for a statistically
meaningful relationship between between ULX number, position, average
luminosity and local metallicity will provide a definitive test of our
proposal.

\section{Acknowledgements}

LZ acknowldges financial support from INAF through grant PRIN-2007-26.
We thank the referee for constructive comments that helped to improve
a previous version of this paper.

%%%%%%%%%%%%%%%%%%%%%%%%%%%%%%%%%%%%%%%%%%%%%%%%%%%%%%%%%%%%%%%%%%%%%%%%%%%%%
%
%
%
%

%
%
%
%
%%%%%%%%%%%%%%%%%%%%%%%%%%%%%%%%%%%%%%%%%%%%%%%%%%%%%%%%%%%%%%%%%%%%%%%%%%%%%
%
%
%
%
\end{document}